\documentclass{WileyChemistry-template}

\usepackage{graphicx}
\usepackage{dcolumn}
\usepackage{bm}

\usepackage[utf8]{inputenc}
\usepackage[T1]{fontenc}
\usepackage{mathptmx}
\usepackage{xcolor}
\usepackage{layouts}
\usepackage{graphicx}
\usepackage{caption}
\usepackage{subcaption}
\usepackage{tabularray}
\usepackage{comment}
\usepackage{ulem}
\usepackage{amsmath}
\usepackage{lipsum}
\usepackage{multirow}
\usepackage{physics}
\usepackage{braket}

\definecolor{zgreen}{rgb}{0.0, 0.26, 0.15}
\newcommand{\quasi}[1]{\textit{quasi}}
\long\def\comment#1{}

\usepackage{booktabs}

\title{\raggedright Simplification of  the Fermi-L\"owdin Self-Interaction Correction Method for Efficient 
Self-Interaction-Free  Density Functional Calculations }

\author{
\begin{minipage}{\textwidth}
	Selim Romero,\textsuperscript{[a]} Yoh Yamamoto,\textsuperscript{[b]} Tunna Baruah,\textsuperscript{[a,b]} Rajendra R. Zope\textsuperscript{[a,b]} 
\end{minipage}
}

\newcommand{\affiliation}{
\begin{itemize}

\item[{[a]}] S. Romero, Dr. T. Baruah, Dr. R. R. Zope*\\
Computational Science Program, The University of Texas at El Paso, El Paso, Texas 79968\\
E-mail: rzope@utep.edu

\item[{[b]}] Dr. Y. Yamamoto, Dr. T. Baruah, Dr. R. R. Zope\\
Department of Physics, The University of Texas at El Paso,TX, 79968\\

\end{itemize}
}

\renewcommand{\abstract}{
     Fermi-L\"owdin orbital self-interaction-correction 
     (FLOSIC) method uses symmetric orthogonalized 
    Fermi orbitals as localized orbitals in one-electron SIC schemes. In FLOSIC,
    a set of Fermi orbital descriptors (FOD) that define the FLOs is obtained by 
    energy minimization. Determination of optimal FODs is 
    a computationally very demanding task. 
    Here, we propose to simplify the FLOSIC calculations 
     by removing 
    self-interaction error (SIE) from a set of selected orbitals of interest (SOSIC). We illustrate the approach by choosing a set of valence 
    orbitals as active orbitals in SOSIC. 
    The results of a wide range of properties obtained using the valence SOSIC (vSOSIC)  scheme are compared with those obtained with the Perdew-Zunger SIC.
    The two methods agree within a few percent for the majority of the properties.
    The mean absolute error in the vertical detachment energy 
    of water cluster anions  with 
    vSOSIC-PBE against 
    benchmark CCSD(T) results is only 15 meV making vSOSIC-PBE an excellent 
    alternative to the CCSD(T) for the case. 
    The calculation on the [Cu$_2$Cl$_6$]$^{2-}$ complex demonstrates that 
    the FOD optimization in vSOSIC is 
    substantially smoother and faster.
    Assessment of the performance of SIC-r$^2$SCAN shows that it performs similarly to the SIC-SCAN for most properties, but for atomization energies, SIC-r$^2$SCAN outperforms SIC-SCAN. 
}

\newcommand{\keywords}{
	Electronic structure \textbullet\ 
	Density functional calculations \textbullet\ 
	Quantum mechanics \textbullet\ 
	Self-interaction correction 
}

\begin{document}

\twocolumn[\vspace{-1.5cm}\maketitle\vspace{-1cm}
	\textit{\dedication}\vspace{0.4cm}]
\small{\begin{shaded}
		\noindent\abstract
	\end{shaded}
}

\begin{figure} [!b]
\begin{minipage}[t]{\columnwidth}{\rule{\columnwidth}{1pt}\footnotesize{\textsf{\affiliation}}}\end{minipage}
\end{figure}

\section{Introduction}
 
The low computational expense combined with relatively good accuracy of density functional theory (DFT)\cite{PhysRev.140.A1133,levy1979universal}
has made it  a quantum mechanical method of choice to 
study the electronic structure of various types of materials, from atoms and molecules to 
nanostructures to periodic materials.  
Practical DFT calculations require approximation to the unknown exchange-correlation functional, and  numerous density functional approximations (DFAs) 
with varying degrees of complexity 
have been proposed. Many failures of the DFAs have been ascribed to the self-interaction error (SIE) present
in the approximate exchange-correlation functionals. 
The problem  arises since the self-Coulomb energy is not completely canceled by 
the self-exchange energy when the exact, but unknown, exchange-correlation functional is approximated.
A few illustrative  examples of the failures of DFA are charge delocalization in proteins,\cite{fouda2016does} 
completely different charge distribution on Kevan structure for the solvated electron,\cite{WOS:000327712800019}
spurious charge transfer in organic acid-base co-crystals,\cite{leblanc2018pervasive} 
severe overestimation of hyperpolarizabilities in conjugated molecules,\cite{autschbach2014delocalization}
structural distortion in the electron polaron model systems,\cite{rana2022detection}
a lack of size-intensivity of ionization potential\cite{sosa2015size}  
etc. 
The self-interaction correction (SIC) methods to remove SIE in an orbital-wise manner were devised long ago.\cite{
lindgren1971statistical,
perdew1979orbital,perdew1981self,
lundin2001novel,PhysRevA.15.213,zunger1980self,gunnarsson1981self,manoli1988generalized,guo1991alternative}

The most well-known one-electron SIC method is the 
Perdew-Zunger SIC  (PZSIC)\cite{perdew1981self,zunger1980self} method wherein
an orbital by orbital correction is applied to the DFA total energy. \cite{perdew1981self}
The  PZSIC energy is given by
\begin{equation}
        E^{PZSIC}[\rho_\uparrow,\rho_\downarrow] = E^{DFA}[\rho_\uparrow,\rho_\downarrow] -  \sum_{i\sigma}^{occ}\{ U[\rho_{i\sigma}] + E_{XC}^{DFA}[\rho_{i\sigma},0] \} .
\end{equation}
        Here,  $E^{DFA}$ is the DFA total energy, $\rho_\uparrow$, $ \rho_\downarrow$, and $\rho_{i\sigma}$ are up spin, down spin, and orbital densities, respectively;
        $U[\rho_{i\sigma}]$  and  $E_{XC}^{DFA}[\rho_{i\sigma},0] $  are the Coulomb and approximate exchange energies.
   The PZSIC energy minimization corresponds to finding an optimal unitary transformation 
of canonical Kohn-Sham orbitals and results in a 
set of $M(M-1)/2$  conditions (for $M$ occupied orbitals) known as the localization equations\cite{doi:10.1063/1.446959,doi:10.1063/1.448266} 
or symmetry conditions.\cite{legrand2002comparison,messud2008improved}
These equations are given by 
\begin{align}    
 \langle \phi_i \vert V_i - V_j  \vert \phi_j \rangle =  0. 
 \end{align}
 Here,  $V_i$ is the sum of Coulomb and exchange-correlation potential 
 of the $i^{th}$ orbital. Although not as popular as standard gradient-based DFAs, several researchers have adopted PZSIC. \cite{gunnarsson1981self,harrison1983self,harrison1983improved,Heaton_1984,heaton1980electronic,gudmundsdottir2015calculations,jonsson2015towards,lehtola2016effect,klupfel2012effect,lehtola2014variational,vydrov2005ionization,vydrov2004effect,lehtola2016complex,janesko2022systematically,bylaska2006new,ruzsinszky2007density,ruzsinszky2006spurious,jackson2019towards,patchkovskii2002improving,kummel2003two,korzdorfer2008self,baruah1994positron,polo2002electron,heaton1983self,goedecker1997critical,svane1996electronic,svane2000rare,rieger1995self,zope1999atomic,hamada1986self,biagini1994self,xie1999obtaining,arai1995electronic,stengel2008self,price1999application, legrand2002comparison, messud2008improved,polo2002electron,pietezak2021n}
 Many implementations of the PZSIC use localized orbitals obtained using various criteria.\cite{doi:10.1063/1.481421,PhysRevB.75.045101,patchkovskii2002improving}
 In 1984, Luken and Culberson observed that properties of the Fermi hole can be used to transform canonical orbitals into a set of 
 localized orbitals.\cite{Luken1982,Luken1984}
 As these orbitals are not orthogonal, they proposed a symmetric orthogonalization procedure to obtain a set of 
 orthogonal orbitals.
 In 2014, Pederson, Ruszinsky, and Perdew used these orbitals, which they called Fermi-L\"owdin orbitals, to obtain PZSIC 
 energy.\cite{doi:10.1063/1.4869581}  This results in a unitary invariant implementation of the PZSIC energy functional (Eq. 1).  
 The Fermi orbitals (FO)\cite{Luken1982,Luken1984} are  given by 
\begin{equation}
        F_{j\sigma}(\vec{r}) = \frac{\sum_i \psi_{i\sigma}(\vec{a}_{j\sigma}) \psi_{i\sigma}(\vec{r})}{\sqrt{\rho_{\sigma} (\vec{a}_{j\sigma})}} .
\end{equation}
Here, the sum of $i$ represents the sum over Kohn-Sham orbitals ($\psi_{i\sigma}$) and $j$ is the FO index (local orbital),  
$\rho_\sigma$ is the total electron spin density, and  $\vec{a}_{j}$ is the so-called Fermi orbital descriptor (FOD) position.
   The Fermi orbitals are further orthogonalized using the L\"owdin method to give the Fermi-L\"owdin 
   orbitals (FLOs). The PZSIC energy is minimized  
   by varying the FOD positions (parameters)   in combination with 
  a conjugate gradient or BFGS algorithm.\cite{PEDERSON2015153,doi:10.1063/1.4907592} 
   The FLOSIC method\cite{doi:10.1063/1.4869581,doi:10.1063/1.4907592,PEDERSON2015153,yang2017flojac,diaz2021kli,diaz2021real}  ensures size extensivity, as well as unitary invariance of the total energy. 
   It also simplifies the problem since instead of the satisfaction of    $N(N-1)/2$ equations only $3N$ variables, where $N$ is the number of orbitals,
   need to be optimized.  
   This results in a significant reduction in the formal cost of SIC calculations.
   
      The FLOSIC method has been used to study 
   a wide range of chemical and physical properties.\cite{ruzsinszky2007density,perdew1981self,goedecker1997critical,vydrov2006scaling,PERDEW20151,goedecker1997critical,vydrov2006scaling,csonka1998inclusion,johnson2019dipoles,diaz2021implementation,bhattarai2021exploring,schwalbe2020pyflosic,yamamoto2020assessing,yamamoto2020improvements,akter2021static,akter2021how,withanage2019dipole,akter2020dipole,joshi2018fermi,mishra2022coupling,mishra2022barrier,romero2023spin,li2020application,karanovich2021electronic,schwalbe2018ips,aquino2020fractional,yamamoto2019scan,withanage2018question,kao2017opt,withanage2021properties,diaz2021dcep,nguyen2021quick,schwalbe2019interpretation,trepte2021bonding,wagle2021waterion,sharkas2020water,vargas2020waterion,jackson2019towards,sharkas2018shrinking,trepte2019grad,santra2019how,shahi2019stretched} 
   As mentioned above, obtaining the SIC energy in the PZSIC requires the determination of the optimal FOD positions.
   In practice, however, the optimization of FOD is a slow process due to the complicated/shallow potential energy surface generated by the FODs, especially for systems containing 
   transition metal (TM) atoms. The FOD optimization of systems with TM atoms is particularly difficult and can often require a few hundred steps. Moreover,  the number of
   steps required grows as the number of FODs increases. 
   Our experience with the FOD optimizations shows that
   the difficulties primarily arise from the optimization of core FODs.   To alleviate this problem,
   approaches such 
   as freezing the core FOD for the 1s orbitals or the use of pseudopotential have been adopted in 
   some FLOSIC calculations on 
   the transition metal and some organic 
   complexes.\cite{joshi2018fermi,karanovich2021electronic,ruan2023spin}
    
   In this work, we propose a simplification of the FLOSIC method for more efficient FLOSIC
   calculation. This is done by selecting a subset of Kohn-Sham orbitals to construct FLOs for the SIC calculations. 
   Since the physical properties of systems are determined mainly by the valence electrons,
   we choose this subset to be the valence orbitals.
   Alternatively, one can also choose it to be the core orbitals if interested in studying core excitation.
   We validate  this approach 
   by performing extensive tests on a variety of systems on several different properties. We show 
   that the proposed simplified scheme,  which results in a substantial reduction in the number of 
   parameters (FODs) to be optimized, reproduces  results of the  \textit{full} all-electron FLOSIC method within 1-2 kcal/mol and 
   that for many properties, in fact,
   provides slightly  improved results. 
   We have applied the present approach to the lowest three rungs of functionals, namely, local spin density approximation (LSDA), Perdew-Burke-Ernzerhof (PBE) generalized gradient approximation (GGA), and r$^2$SCAN meta-GGA functionals. 
   Though not the primary focus of this work, we also report the assessment of the SIC-r$^2$SCAN approach.
   We believe this is the first time SIC-r$^2$SCAN approach has been assessed for a range of electronic 
   properties.
   We also demonstrate that accurate estimates of the vertical detachment energies of water cluster anions can be obtained using the simplest version of the SOSIC, in which only the orbital containing the extra electron in the water cluster anions is corrected for the SIE.

   The details about the present approach and its implementation are provided in the next section, followed by results and discussion.

\section{Methodology}
 We divide the $N$ occupied Kohn-Sham (KS) orbitals into groups of $P$ passive and $(N-P)$ active orbitals  such that  
 the corrected exchange-correlation energy can be written as \cite{yamamoto2020improvements} 
 \begin{equation}\label{eq:sosic-dfa}
   \begin{split}
   E_{XC}^{SOSIC-DFA}  = & E_{XC}^{DFA}[\rho_\uparrow,\rho_\downarrow]
    -\sum_{i\sigma=1}^{P}X_{i\sigma}^{k}\left(U[\rho_{i\sigma}]+E_{XC}^{DFA}[\rho_{i\sigma},0]  \right) \\
    & -\sum_{i\sigma=P+1}^{occ}Y_{i\sigma}^{k}\left(U[\rho_{i\sigma}]+E_{XC}^{DFA}[\rho_{i\sigma},0]  \right).
    \end{split}
\end{equation}
Here, $X_{i\sigma}^{k}$ and $Y_{i\sigma}^{k}$ can be considered as scaling factors that can be determined using various 
criteria.\cite{vydrov2006scaling,yamamoto2020improvements}
The set of $P$ passive orbitals should be chosen carefully. For example, by applying full SIC correction to the orbitals that 
 participate in  stretched bonds and scaling down the SIC for other orbitals, Yamamoto and coworkers 
were able to obtain barrier heights of BH6 dataset within the chemical accuracy\cite{yamamoto2020improvements} 
{using this selective orbital scaling SIC (SOSIC)}.
As our purpose in this work is to simplify the SIC approach to improve its computational efficiency,
we choose $P$ to be the core orbitals with factors $X^k_{i\sigma}=0$ and $Y^k_{i\sigma}=1$. 
This amounts to removing the SIE only from  the valence electrons. Other choices of $P$ are possible but here
we choose to select only the valence electrons so that direct comparison of the results with earlier
PZSIC results  is possible and meaningful.
This method of selectively applying SIC to the valence electron is referred to as vSOSIC hereafter.
Likewise, the approach that uses the FLOs constructed from the core KS orbitals 
( $X^k_{i\sigma}=1$ and $Y^k_{i\sigma}=0$ in Eq. [\ref{eq:sosic-dfa}] ) to compute
the SIC of the core orbitals will be called cSOSIC.

We choose  the $P$ core electrons   as  shown in Table \ref{tab:ecp_shell_removal}. 
For example, in the case of manganese which has an electronic configuration [Ar]$4s^2 3d^{5}$,
$P$ is $10$, i.e., only the  electrons in the $3s, 3p, 3d$ and  $4s$ are considered for the SIC calculations. 
  Previous calculations indicate that a full shell should be included to allow shell hybridization in 
the SIC calculations. It is noted that the FOD positions for atoms often follow the shell structure such that four FODs transcribe to the second shell ($2s$,$2p$), 9 for the third shell ($3s$,$3p$,$3d$), etc. Therefore, the use of $[Ar]$ core, in this case, is not recommended.
 The 3d transition metal atoms from scandium to zinc are assigned  a neon core similar to the small core in
 the effective core potential (SC ECP) schemes. 
Such a choice considerably simplifies SIC calculations due to  reduction in 
the time-consuming task of calculating orbitalwise SIC potentials. 
More importantly, it also facilitates the optimization 
of the FODs since FODs in all orbital FLOSIC calculations are often more difficult to optimize.
We note
in passing that the proposed scheme 
            also permits the application of SIC in selected regions in  space in the spirit of embedding approaches
for applications such as single-atom catalysis. Such applications will be pursued in subsequent studies. 

In the vSOSIC method, localized Fermi orbitals are constructed from the valence 
Kohn-Sham orbitals and valence electron density as,
\begin{equation}\label{eq:sosic-flo}
        F_{j\sigma}(\vec{r}) = \frac{\sum\limits_{i=P+1}^{N} \psi_{i\sigma}(\vec{a}_{j\sigma}) \psi_{i\sigma}(\vec{r})}{\sqrt{\rho^{val}_{\sigma} (\vec{a}_{j\sigma})}} .
\end{equation}

\begin{figure}
    \centering
    \includegraphics[width=0.7\columnwidth]{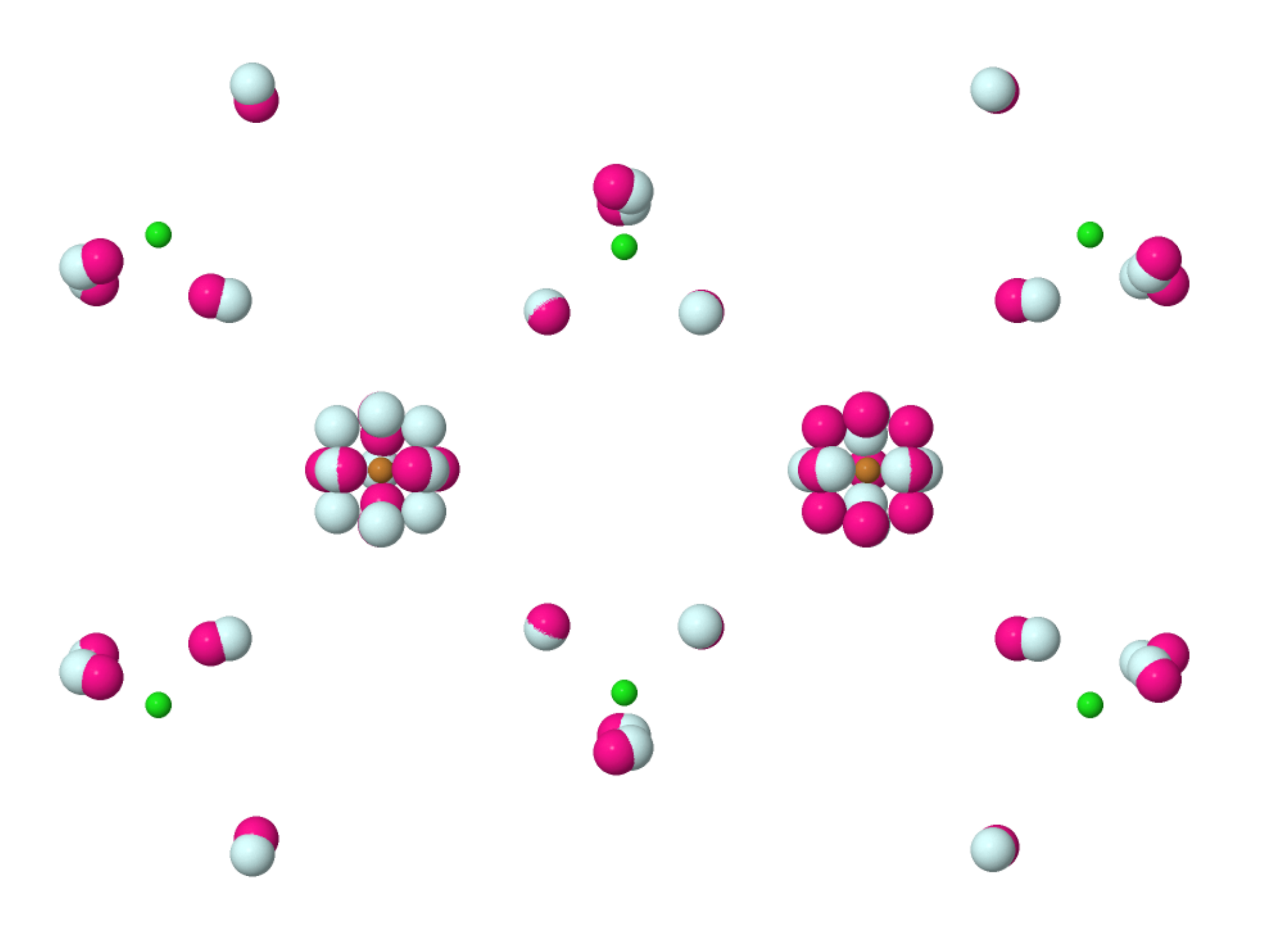}
    \caption{[Cu$_2$Cl$_6$]$^{2-}$  SOSIC Fermi orbital descriptors. The magenta (grey) FOD corresponds spin-up (down) channel. The orange (green) dots represent copper (chlorine) atom positions. Instead of 162 FODs, vSOSIC uses only 82 FODs.}
    \label{fig:cu2cl6_fod}
\end{figure}

Here, $ \rho^{val}_{\sigma} (\vec{a}_{j\sigma}) =
\sum_{i=P+1}^{N} |\psi_{i\sigma}(\vec{a}_{j\sigma}) |^2 $,
$N$ is the number of electrons, and $\vec{a}_{j\sigma}$ are the Fermi-orbital descriptors. The Fermi-L\"owdin orbitals  ($\phi_{i\sigma}(\vec {r}) $) are obtained after symmetric orthogonalization of the Fermi orbitals from Eq. (\ref{eq:sosic-flo}).
These FLOs are used to compute the SIC potentials and energies.
Naturally, vSOSIC requires less number of FODs than the number of occupied orbitals. FODs for [Cu$_2$Cl$_6$]$^{2-}$ are shown in Fig.~\ref{fig:cu2cl6_fod} as an example where 82 FODs are present instead of the 162 electrons present in the molecule.

\begin{figure}
    \centering
    \includegraphics[width=\columnwidth]{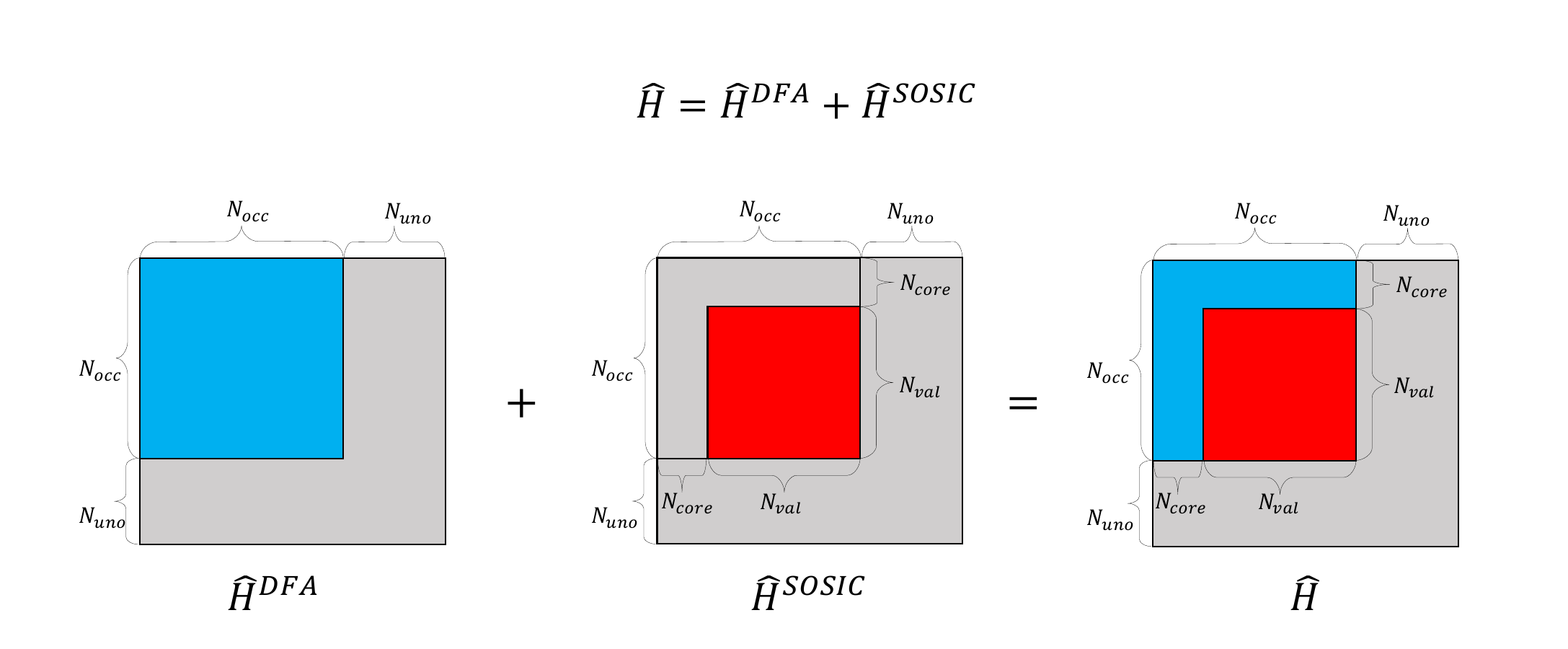}
    \caption{Visual representation of the DFA+SIC Hamiltonian construction in the vSOSIC approach.}
    \label{fig:sosic-ham}
\end{figure}
\begin{figure}
    \centering
    \includegraphics[width=0.8\columnwidth]{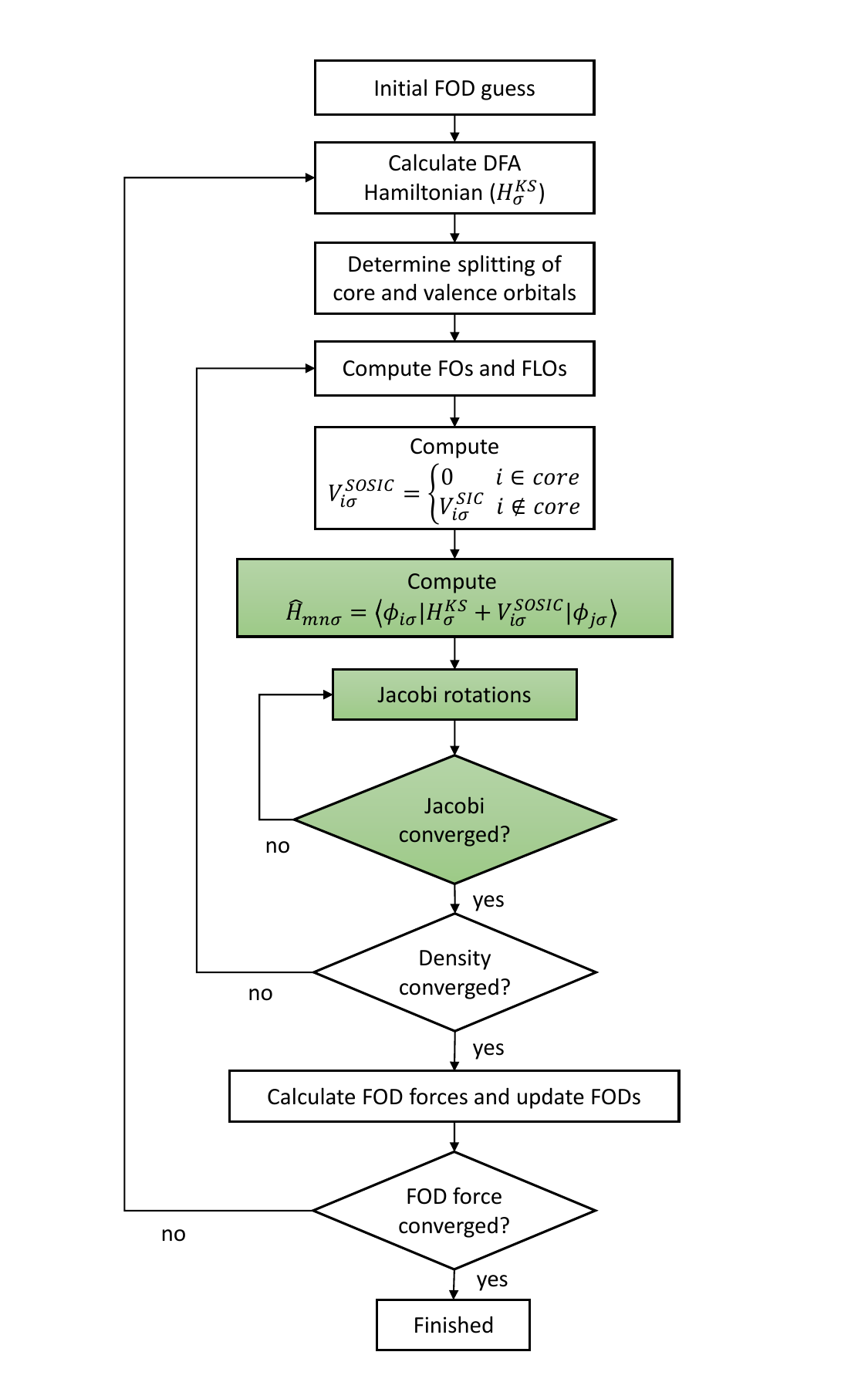}
    \caption{vSOSIC algorithm flowchart for Jacobi rotations approach.}
    \label{fig:jacobi-flowchart}
\end{figure}

Self-consistency  can be 
obtained either using optimized effective potential within the Kriger-Li-Ifarate approximation \cite{diaz2021kli}
or  using the Jacobi update approach.\cite{yang2017flojac} We have used the Jacobi update approach in this work.   We first construct a SIC Hamiltonian 
as described by Yang \textit{et al.} \cite{yang2017flojac} as  
\begin{equation}\label{eq:sosic-ham}
       H_{\sigma} = H^{DFA}_{\sigma}+\sum_{i,j=P+1}^{N_{\sigma}}\frac{1}{2} (V_{ij}^{i\sigma}+V_{ji}^{j\sigma}) \ket{\phi_{i\sigma}}  \bra{\phi_{j\sigma }} 
\end{equation}
where 
$$V_{ij}^{i\sigma}= \mel{\phi_{i\sigma}}{V^{i\sigma}}{\phi_{j\sigma}}$$
with $V^{i\sigma}$ being the SIC potential for the $i$th orbital of spin $\sigma$. The Hamiltonian matrix is constructed in the basis of the canonical orbitals. Fig. {\ref{fig:sosic-ham}} pictorially presents how the vSOSIC Hamiltonian is constructed. For the vSOSIC case, the SIC part of the  Hamiltonian has   $M\times M$ non-vanishing elements where $M=N-P$.
The Jacobi update approach is used to derive the orthogonal eigenvectors. Once the self-consistency is achieved for a given FOD configuration,
the forces on the FODs are calculated \cite{doi:10.1063/1.4907592,PEDERSON2015153}and the FOD positions are optimized either using the LBFGS or conjugate gradient schemes.\cite{Liu1989}  We show in Fig. \ref{fig:jacobi-flowchart} the vSOSIC algorithm with Jacobi approach.

The vSOSIC calculations are performed for the \textit{non-empirical} functionals at the three lowest rungs 
of functional ladder. These are LSDA, GGA, and meta-GGA. We choose PW92 correlation functional\cite{PhysRevB.46.6671} for the LSDA,
PBE parameterization for the GGA,\cite{PhysRevLett.77.3865,PhysRevLett.78.1396} and r$^2$SCAN meta-GGA functionals.\cite{furness2020r2scan}
For comparison, we have also included PZSIC calculations using FLOs where FLOs are constructed 
using all Kohn-Sham orbitals as in previous FLOSIC calculations. The PZSIC-LSDA and PZSIC-PBE 
calculations have been reported earlier while the PZSIC-r$^2$SCAN results reported herein 
are new results.
\comment{
Earlier studies have shown that the overcorrecting tendency of the PZSIC can be mitigated by scaling
the SIC energy. 
We scaled SOSIC by a suitable constant factor to improve the performance of HOMO eigenvalues when compared to experimental ionization potentials (IPs) of 37 selected molecules. A more detailed explanation is provided in Subsection \ref{sec:homo-evals}.}

We have modified  the FLOSIC code \cite{FLOSICcode} for the work described here. The NRLMOL basis set is used for all calculations in this work.\cite{PhysRevA.60.2840}
For the vertical detachment energy calculations of water cluster anions, 
we used the NRLMOL basis set\cite{PhysRevA.60.2840} with extra diffused functions  to account for the anionic nature of the water clusters.\cite{herbert2005calculation}
The Gaussian exponents for the extra functions are provided by Yagi \textit{et al.}
\cite{doi:10.1021/jp802927d} where their values are $9.87\times 10^{-3}$, $8.57\times 10^{-3}$, and $3.72\times 10^{-3}$ for oxygen s and p, and hydrogen s functions, respectively. 

As mentioned earlier this is the first work that reports 
FLOSIC-r$^2$SCAN calculations on a wider range of properties. We therefore briefly 
comment on the numerical details of SIC calculations with r$^2$SCAN. In our earlier works,\cite{yamamoto2019scan,yamamoto2020assessing}
we have implemented and discussed the performance and numerical sensitivity
of the SCAN and rSCAN functionals.
The sensitivity of the SCAN functional to the choice of the numerical grid has 
been noted in a few studies. \cite{doumont2022implementation,doi:10.1063/1.5094646,doi:10.1063/1.5120408,doi:10.1063/1.5128484,yamamoto2019scan,yamamoto2020assessing}
The numerical instability of SCAN is primarily due to its interpolation function, which is smoothed out 
in rSCAN and r$^2$SCAN functionals. 
The r$^2$SCAN functional\cite{furness2020r2scan} is numerically more stable than the original SCAN and requires a less
dense grid. We note that SIC calculations typically require denser numerical grids
than the standard DFA calculations as the SIC contributions to the Hamiltonian/Fock 
matrix elements and SIC energy corrections are evaluated using the orbital densities 
that can vary far more rapidly than the total spin densities used in evaluating
corresponding DFA contributions.
Our earlier work has shown that FLOSIC calculations with 
SCAN functional typically requires very dense grids with about 140000 grid points
per atom. We have examined the numerical needs of r$^2$SCAN in the FLOSIC calculations.
The details are in  the supplementary information. We find that FLOSIC-r$^2$SCAN calculations
require a factor of 2-4 times fewer grid points than FLOSIC-SCAN calculations.
To describe the energy landscape in covalent bond stretching or reaction pathway, however, it requires a denser mesh than the default mesh of the FLOSIC code. 
The numerical mesh used in the present SIC-r$^2$SCAN calculations  
has roughly 1.1--1.5 times more grid points than the mesh requirements of the SIC-LSDA.

\begin{table*}
\caption{\label{tab:ecp_shell_removal} 
Sets of criteria for determining the value $P$ in Eq. (\ref{eq:sosic-dfa}) 
for hydrogen up to the barium atom in order to differentiate core and valence electrons 
similar to the criteria used in the ECP approach discussed in Ref. \citenum{hay1985ab}.}
\begin{tabular}{cclcl}
\toprule
    & \multicolumn{2}{c}{Criteria \#1}		&	\multicolumn{2}{c}{Criteria \#2}	\\
     & \multicolumn{2}{c}{Large core (LC)}		&	\multicolumn{2}{c}{Small core (SC)}	\\   
    \cmidrule(lr){2-3} \cmidrule(lr){4-5}

$Z$	&	No. core	&	core-shell   &   No. core    &   core-shell	\\ \hline
1-4	    &	0	&	  &   0 &        	\\
5-12	&	2	&	He: $1s^2$   &  2 &	He: $1s^2$	\\
13-30	&	10	&	Ne: [He] $2s^{2} 2p^6$ &  10 & Ne: [He] $2s^{2} 2p^6$  	\\
31-48	&	28	&       [Ar] $3d^{10}$ &  10 & Ne: [He] $2s^{2} 2p^6$  	\\
49-56	&	46	&	[Kr] $4d^{10}$  &   28	& [Ar] $3d^{10}$	\\
\bottomrule
\end{tabular}
\end{table*}

\section{Results and discussion}

In this section, we first present the results on energy-related properties such as  
atomization energies,
reaction barrier heights,  ionization potentials, and magnetic exchange coupling parameters 
using the present vSOSIC approach to assess
its performance in comparison to the \textit{all-orbital} PZSIC  
results
in which all canonical KS orbitals
are included in the construction of the Fermi-L\"owdin orbitals and SIC energy is calculated by summing
SIC energy contribution of \textit{all} orbitals.

\subsection{Atomization energies}\label{sec:atomization}

\begin{sloppypar}
We evaluated the performance of vSOSIC atomization energies (AEs) on the AE6\cite{truhlar2003small} dataset and a set of 37 molecules from the G2/97 dataset.\cite{truhlar2011communication} 
AE6 consists of the atomization energies of six molecules and is typically used as a small representative benchmark set of the larger main group atomization energy (MGAE109) dataset. We calculated atomization energies as follows, $AE =  \sum_i^{N_{atoms}} E_i - E_{mol}$, where $E_i$ is the energy of the atoms and $E_{mol}$ is the energy of the molecule. 
 The mean absolute errors (MAEs) for the AE6 set are derived by comparing against the values from Ref. \citenum{truhlar2003small} and are shown in Table \ref{tab:mae_ae}. 
Since the vSOSIC includes a smaller set of orbitals, we can expect the errors in atomization energies to be between those for DFA and PZSIC-DFA  errors. This trend can be seen in Table \ref{tab:mae_ae}.
\end{sloppypar}

\begin{table*}
\caption{\label{tab:mae_ae} 
Mean absolute error (MAE) in kcal/mol and mean absolute percentage error (MAPE) in \% of atomization energy for the data sets AE6 and 37-molecules.
}

\begin{tabular}{lccccc}
\toprule
Functional & Method   & \multicolumn{2}{c}{MAE (kcal/mol)}		&	\multicolumn{2}{c}{MAPE (\%)}	\\
    \cmidrule(lr){3-4} \cmidrule(lr){5-6}

	&		&	AE6 & 37-molecules	&	AE6  &	37-molecules \\\hline
LSDA & DFA		&	74.3$^\text{[a]}$	&  64.5   &	15.9$^\text{[a]}$ &   24.2$^\text{[b]}$ 	\\
LSDA & PZSIC		&	58.0$^\text{[a]}$ & 46.8$^\text{[b]}$  & 9.4$^\text{[a]}$ & 13.4$^\text{[b]}$ 	\\
LSDA & vSOSIC	&	66.6 &  51.1 &	11.0    &	13.8\\
LSDA & cSOSIC & 99.3 & 65.2 & 21.0  & 24.2 \\
\hline

PBE & DFA		&	13.4$^\text{[a]}$	&	23.7$^\text{[b]}$ &    3.3$^\text{[a]}$     &  8.6$^\text{[b]}$  	\\
PBE & PZSIC		&	18.8$^\text{[a]}$  &   20.2$^\text{[b]}$ 	&	6.8$^\text{[a]}$    &   9.7$^\text{[b]}$  	\\
PBE & vSOSIC		&	16.3  &  22.7 	&	5.3	   & 9.6   \\
PBE & cSOSIC & 34.9 & 23.6 & 6.6 & 8.7\\
\hline

r$^2$SCAN	&	DFA	&	3.0	&	 16.0   &   1.9 & 6.1     	\\
r$^2$SCAN	&	PZSIC	&	26.3    &   16.7	&	6.9	&   9.9 \\
r$^2$SCAN	&	vSOSIC	&	17.2    &   17.1	&	5.1	 & 8.9  \\
r$^2$SCAN & cSOSIC & 25.9 & 15.7 & 5.1 & 5.5\\
\bottomrule
\end{tabular}

\footnotesize{\textsf{[a] Reference \citenum{zope2019step}. 
[b] Reference \citenum{yamamoto2019scan}.}}
\end{table*}

Additionally, we studied the 37  molecules used in our earlier work (Ref. \citenum{yamamoto2019scan}).
We experienced SCF convergence issues with the vSOSIC approach for LiBr and NaBr with Jacobi rotation when a large core SOSIC treatment is used on the bromine atom.
This convergence issue may be because of core-active mixing.
In such cases, using small core SOSIC can eliminate the issues.

The qualitative performance of vSOSIC using a small core for the larger atoms is summarized  in Table \ref{tab:mae_ae} compared against Ref. \citenum{truhlar2011communication}. 
The trends in atomization energies for the AE6 and the larger set are similar for LSDA in that the MAEs are reduced with PZSIC-LSDA compared to DFA-LSDA. With PBE and r$^2$SCAN functionals, the MAEs for the larger set are comparable with the three approaches.   On the other hand, with AE6 set, the application of SIC introduces large errors with r$^2$SCAN. Since AE6 is a small set, the larger set is more likely to show the general performance of these functionals. Overall, the vSOSIC-DFA MAEs are close to those of PZSIC-DFAs. 
For comparison, we include cSOSIC in Table~\ref{tab:mae_ae} where the core-only correction results in rather poor performance for the AE6 set.

\subsection{Barrier heights}

The barrier heights of chemical reactions are difficult to describe correctly with a DFA since the SIE appears in
stretched bond situations at the saddle point calculation. 
 DFAs due to SIEs tend to incorrectly provide lower energies for the transition
states.
Previously, PZSIC performance on the BH76 set was studied\cite{mishra2022study}, and it is reported that both accurate energy functional as well as accurate electron density are important for describing barrier height calculations.
Furthermore, previous SOSIC work\cite{yamamoto2020improvements} on the reaction barrier heights of the BH6 dataset\cite{truhlar2003small}
showed that SOSIC works well when it includes only the orbitals corresponding to 
the stretched bonds.

Here, we systematically compare the performance of the vSOSIC methods on reaction barrier heights using the BH6 set.
The reactions of BH6 are (a) OH + CH$_4$ $\rightarrow$ CH$_3$ + H$_2$O,  (b) H + OH $\rightarrow$ H$_2$ + O, and (c) H + H$_2$S  $\rightarrow$ H$_2$ + HS.
There are six barrier heights from the combined forward and reverse reaction pathways.
Table \ref{tab:mae_bh} provides a summary of the results and
shows that vSOSIC performance is nearly identical to that of  PZSIC with comparable MAEs for the three functionals.

\begin{table}
\caption{\label{tab:mae_bh} 
Mean absolute error (MAE) in kcal/mol for the reaction barrier heights of BH6 and WCPT18 sets. 
}
\begin{tabular}{lccc}
\toprule
Functional 	&	Method   	&	\multicolumn{2}{c}{MAE (kcal/mol)}	\\
    \cmidrule(lr){3-4} 
    &		 &	BH6   &    WCPT18 \\\hline
     
LSDA	&	DFA	&17.6$^\text{[a]}$   &   17.7  	\\
LSDA	&	PZSIC	&4.9$^\text{[a]}$    &	7.5   \\
LSDA	&	vSOSIC	&5.2  &   7.6	\\\hline
PBE	&	DFA	&8.0$^\text{[a]}$  &   8.8  \\
PBE	&	PZSIC	&4.2$^\text{[a]}$   &   10.3  	\\
PBE	&	vSOSIC	&4.1    &   9.4	\\\hline
r$^2$SCAN	&	DFA	&7.6	 &  6.2 \\
r$^2$SCAN	&	PZSIC	&2.8 &    6.0   	\\
r$^2$SCAN	&	vSOSIC	&2.6	 &   5.5 \\
\bottomrule
\end{tabular}

\footnotesize{\textsf{[a] Reference \citenum{zope2019step}.}}
\end{table}

Since the BH6 set is a rather small set of reaction barriers, we additionally used the WCPT18 set\cite{karton2012assessment} to 
evaluate the vSOSIC performance 
in comparison to PZSIC. 
The WCPT18 set consists of 18 reaction barrier heights and requires 28 single-point calculations.
The set consists of 9 water-catalyzed proton-transfer reactions that involve zero, one, or two water molecules as a catalyst. 
The MAEs of the WCPT18 set  presented in Table \ref{tab:mae_bh} for LSDA, PBE and r$^2$SCAN show that the 
vSOSIC and PZSIC results agree within 1 kcal/mol.
This is expected since the proton transfer reactions involve stretching electron density on the hydrogen atom where SIC is truly needed.
This stretching happens on valence orbitals, where vSOSIC and PZSIC have the same SIC effect on these orbitals.

\subsection{Highest occupied  molecular orbital eigenvalues for 38 selected molecules}\label{sec:homo-evals}

Janak's theorem relates that the negative of the highest occupied molecular orbital (HOMO) eigenvalue in DFT is equivalent to the vertical ionization 
potential (IP).\cite{janak1978,PhysRevA.30.2745,PhysRevB.60.4545, perdew1997comment}
Since many DFAs do not have a correct asymptotic potential character, their exchange-correlation potentials tend to be too shallow in the asymptotic 
region, resulting  thereby in absolute HOMO eigenvalues that underestimate the IPs.
PZSIC is shown to lower the HOMO energy levels by deepening the exchange-correlation potential and widening the HOMO-LUMO gaps. This usually results in a much-improved agreement between absolute of HOMO eigenvalues with experimental IPs or higher-level theories.

\begin{figure}
\centering
\includegraphics[width=\columnwidth]{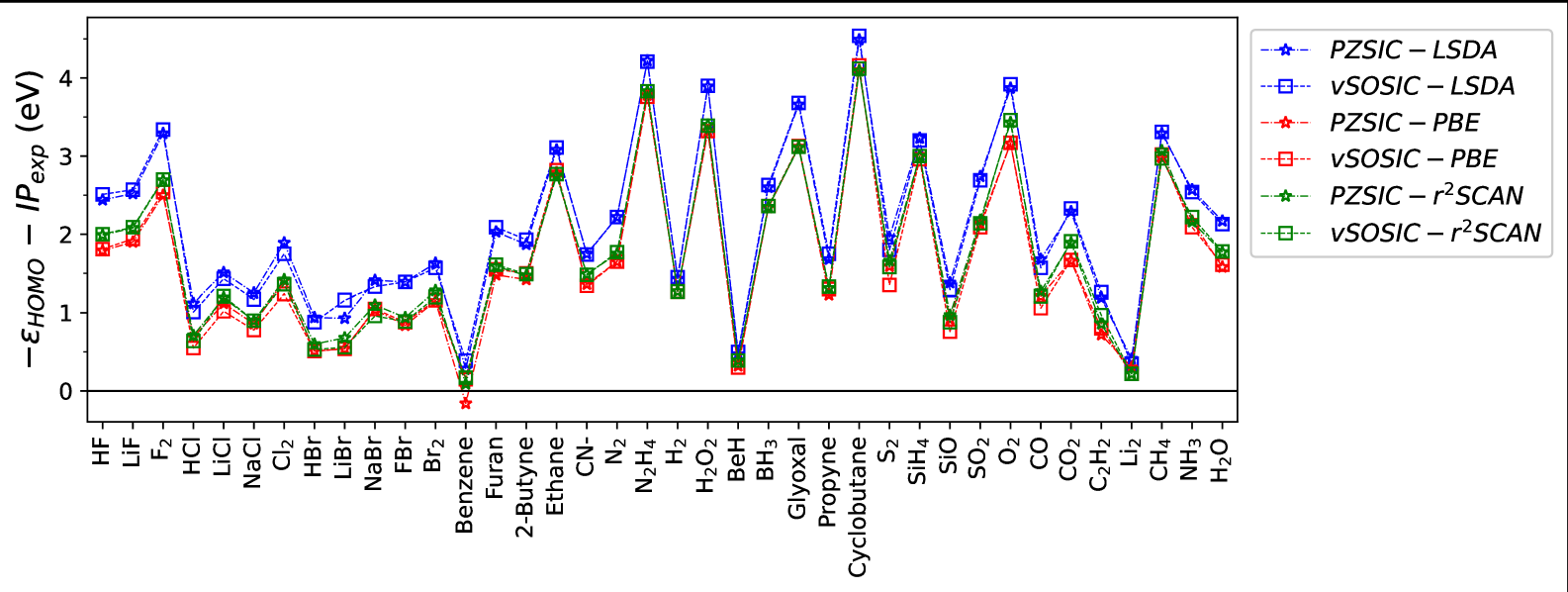}
\caption{Negative of HOMO eigenvalues of 38 selected molecules, a subset of G2/97 set, compared against experimental ionization potential for three functionals and two SIC methods.
}\label{fig:homo-errors}
\end{figure}

Fig. \ref{fig:homo-errors} compares the difference of absolute of HOMO eigenvalues of the 38 molecules 
that are a subset of G2/97 set and  experimental IP for PZSIC-DFA and vSOSIC-DFA for LSDA, PBE, and r$^2$SCAN. 
Although absolute HOMO eigenvalues in PZSIC estimate IPs better than DFA,  
the absolute HOMO eigenvalues of PZSIC overestimate the experimental IPs by up to 4 eV for the set of molecules studied here. 
vSOSIC HOMO eigenvalues are in good agreement with PZSIC.
The eigenvalue spectra of LiBr and NaBr molecules are shown in
Supplementary Materials
where it can be seen that SOSIC and PZSIC valence eigenvalues 
are essentially identical.

\subsection{Vertical detachment energy of water cluster anions}
Electron hydration is an important phenomenon in biological processes.\cite{alizadeh2012precursors,herbert2017hydrated}
Modeling such behavior is challenging for the local and semi-local DFA. The  semi-local GGA and meta-GGA and B3LYP performance 
are rather poor for such systems as the electron attached to the water cluster is delocalized over the system. Typically,
the extent of the delocalization worsens with increasing water cluster size, and consequently,
 post-Hartree-Fock methods are often the methods of choice. 
Recent studies by Vargas and coworkers showed that SIC methods can be 
a good alternative to M{\o}ller-Plesset second-order perturbation theory (MP2) and LC-BOP level of theory for describing the systems of hydrated electrons.\cite{vargas2020waterion}
An excess electron in water clusters can be dipole-bound, trapped on a cluster surface, solvated internally to a cluster, or bound to dangling OH bonds of a cluster.\cite{doi:10.1021/jp802927d} In all situations, active orbitals play an important role.
Earlier studies carried out by this group on electron binding to water clusters have shown that 
PZSIC-PBE can predict the vertical detachment energies (VDE) 
in comparable accuracy as the CCSD(T) level of theory 
when the negative of the HOMO eigenvalue is used to estimate VDE.
To assess the present SOSIC approach, we have computed the vertical detachment 
energies of water clusters from the absolute of the HOMO eigenvalues with 
the PBE and r$^2$SCAN 
functional.  We consider water dimer, five trimers (3AA, 3D, 3I-1, 3I-2, and 3L), four tetramers (4AA, 4D, 4I, and 4L), five pentamers (5AA-1, 5AA-2, 5D, 5I, and 5L), and five hexamers (6AA-1, 6AA-2, 6D, 6I, and 6L).
The same isomer notation is used as in the references \citenum{doi:10.1021/jp802927d,vargas2020waterion}.

The calculated VDEs are shown in Table~\ref{tab:water_ions} where PZSIC-PBE, vSOSIC-PBE, 
vSOSIC-r$^2$SCAN, and CCSD(T) values from Ref. \citenum{doi:10.1021/jp802927d} are compared.
The vSOSIC-PBE performance is very close to PZSIC-PBE  with
MAEs of 16.9 (PZSIC-PBE) and 15.0 meV (vSOSIC-PBE) when compared to CCSD(T) values, respectively. 
The differences between the PZSIC-PBE and vSOSIC-PBE VDE values are 20 meV or less for the majority of the isomers except for 6AA-1 for which the difference is 53 meV.
Similarly, MAE of 42.5 meV for vSOSIC-r$^2$SCAN closely matches MAE of
44 meV found for PZSIC-SCAN in Ref.
\citenum{vargas2020waterion}.

\begin{table*}
\caption{\label{tab:water_ions} 
Vertical detachment energy of water cluster anions is estimated as the absolute values of the HOMO eigenvalues. MAE (in meV) is calculated with respect to CCSD(T) from Reference \citenum{doi:10.1021/jp802927d}.
}
\begin{tabular}{lccccccc}
\toprule
&\multicolumn{3}{c}{PBE} &\multicolumn{2}{c}{r$^2$SCAN} & & \\\cmidrule(lr){2-4}\cmidrule(lr){5-6}
System 	&	PZSIC	&	vSOSIC	&	1orb-SOSIC	& vSOSIC & 1orb-SOSIC &	B3LYP$^\text{[a]}$ & CCSD(T)$^\text{[a]}$	\\\hline
2L      &	30	&	33	&	18	& 41    & 37   &194&	29	\\
3AA     &	180	&	184	&	149	& 211   & 192   &399&	187	\\
3D      &	14	&	17	&	4	& 19    & 16   &184&	6	\\
3I-1    &	205	&	216	&	107	& 253   & 214   &526&	190	\\
3I-2    &	185	&	188	&	146	& 209   & 186   &427&	175	\\
3L      &	148	&	156	&	129	& 177   & 168   &346&	146	\\
4AA     &	314	&	319	&	279	& 362   & 338   &561&	336	\\
4D      &	42	&	62	&	41	& 66    & 59   &239&	49	\\
4I      &	456	&	445	&	314	& 515   & 440   &713&	439	\\
4L      &	236	&	246	&	216	& 273   & 261   &478&	255	\\
5AA-1   &	358	&	385	&	294	& 406   & 373   &600&	370	\\
5AA-2   &	354	&	357	&	315	& 404   & 377   &592&	376	\\
5D      &	73	&	77	&	54	& 81    & 76   &285&	61	\\
5I      &	467	&	473	&	372	& 527   & 462   &757&	469	\\
5L      &	277	&	282	&	254	& 308   & 302   &527&	294	\\
6AA-1   &	521	&	574	&	425	& 619   & 551   &847&	553	\\
6AA-2   &	443	&	456	&	392	& 510   & 475   &706&	477	\\
6D      &	114	&	118	&	87	& 127   & 123   &347&	104	\\
6I      &	904	&	891	&	671	& 966   & 836   &1120&	839	\\
6L      &	358	&	367	&	331	& 253   & 204   &643&	381	\\\hline
MAE	&	16.9	&	15.0&	56.9& 42.5   &	16.9  &238&		\\
\bottomrule
\end{tabular}

\footnotesize{\textsf{[a] Reference \citenum{doi:10.1021/jp802927d}.}}
\end{table*}

Typically, the extra electron in water cluster anions is unbound in the standard DFA calculations with positive HOMO eigenvalues.  As demonstrated above, the vSOSIC-PBE and PZSIC-PBE can accurately describe electron binding in these clusters. 
Thus, the removal of SIE from the valence orbitals can cure the failure of DFAs. 
As an interesting application of vSOSIC, here we further 
examine if removing the SIE for just one orbital (the last orbital with the extra electron) can improve 
the description of electron binding in these clusters. 
This treatment is done self-consistently. 
The results of this one-orbital SOSIC  show that the removal of SIE from  just
one orbital results in electron binding with negative HOMO eigenvalues for all the clusters.
In this case, the VDEs obtained from the HOMO eigenvalues however show a higher MAE of 56.9 meV 
with PBE.
Although this MAE is larger compared to the all orbital PZSIC-PBE, it is still
 comparable to MP2 MAE (44 meV) and far better than that of B3LYP (238 meV).\cite{doi:10.1021/jp802927d} 
The application of 1orb-SOSIC on r$^2$SCAN gives a smaller MAE (16.9 meV) than PZSIC and vSOSIC, however. For the set of systems studied here, the values estimated with 1orb-SOSIC are always smaller than the PZSIC and vSOSIC estimated values. 
It is remarkable that the removal of SIE from just one orbital (extra electron) can result in a major improvement
in the VDEs of water anions since the cost of this calculation is practically the same as that of PBE functional. 
Since the SIC is applied only to the extra electron, the   FLO in this case is same as the Kohn-Sham orbital. This 
approach therefore can be readily introduced in most density functional codes.

\subsection{Magnetic exchange coupling constant of chlorocuprate}
 The magnetic exchange interaction between localized spins is another property what is affected by the delocalization error arising from the SIE in the DFAs.  The 
coupling strength between two magnetic spins is characterized by a quantity known as magnetic exchange coupling constant ($J$), and its sign and magnitude determine the magnetic nature and strength of materials.
The spin Hamiltonian for such interaction is written as
\begin{align}
    H_{spin}= -J \sum_{i,j} \mathbf{S_i} \cdot \mathbf{S_j}.
\end{align}
By relating the DFT energy of high-spin and low-spin states with a given electron configuration to the $H_{spin}$ of the 
corresponding spin configuration, we can determine the  value of $J$ from the DFT calculations.
As an application of the vSOSIC method, we compute  the magnetic exchange coupling constant of [Cu$_2$Cl$_6$]$^{2-}$  system and compare
the results with previous SIC results by Mishra \textit{et al.} and other groups.\cite{mishra2022coupling,joshi2018fermi,MIRALLES1992555,CASTELL1994377} 
With LSDA and PBE, the coupling strength of this molecule is overestimated by a few orders.
Previous PZSIC studies\cite{mishra2022coupling}  using  ECP showed  that both the SIC correction to the 
energy and  to  the density  are needed for accurate descriptions of coupling constants with LSDA and PBE functionals. 
On the other hand, for the SCAN family of functionals, \textbf{sometimes} only SIC correction to the density  may be sufficient.\cite{wagle2021waterion,yamamoto2019scan}

The magnetic exchange coupling constant is computed using the spin projection approach of 
Noodleman\cite{noodleman1981valence} given as a formula, $J=(E_{BS}-E_{HS})/(2S_AS_B)$, 
where S$_A$ and S$_B$ are the spins at two magnetic centers, A and B. $E_{BS}$ is the energy of the molecule in 
broken symmetry spin configuration ($\uparrow\downarrow$) and $E_{HS}$ is energy in high-spin configuration 
($\uparrow\uparrow$). 
The  vSOSIC calculated magnetic exchange coupling constants are compared with previous results  in Table \ref{tab:mag_exg_coup}. 
It is evident that the two methods (PZSIC and vSOSIC) differ at most by $10$ cm$^{-1}$ for the LSDA and PBE functionals. 
We note that the magnetic exchange coupling constant is more sensitive to the accuracy in the total energies than other non-magnetic properties, and it requires tighter convergence criteria in FOD optimizations than usual calculations.  
The density-corrected DFA is a good way to improve the DFA  predictions when DFA  errors are suspected to be due 
to density delocalization errors.
By comparing both DFA@PZSIC-DFA and DFA@vSOSIC-DFA, we find that the difference is 6 and 14 cm$^{-1}$ for LSDA and PBE cases, respectively.

\begin{table}
\caption{\label{tab:mag_exg_coup} 
Magnetic exchange coupling constant $J$ in cm$^{-1}$ for hexachlorocuprate [Cu$_2$Cl$_6$]$^{2-}$ at 
planar $\theta=0^{\circ}$.
}

\begin{tabular}{lccc}
\toprule
Method	&	$J$	\\\hline
LSDA & –354$^\text{[a]}$ \\
PZSIC-LSDA	&	-78$^\text{[a]}$		\\ 
vSOSIC-LSDA	&	-84	\\ 
LSDA@PZSIC-LSDA	&	-131$^\text{[a]}$		\\ 
LSDA@vSOSIC-LSDA	&	-137	\\ 
\hline
PBE & –234$^\text{[a]}$ \\
PZSIC-PBE	&	-94$^\text{[a]}$		\\ 
vSOSIC-PBE	&	-84	\\ 
PBE@PZSIC-PBE	&	-138$^\text{[a]}$		\\ 
PBE@vSOSIC-PBE	&	-124	\\ 
\hline
r$^2$SCAN & –42$^\text{[b]}$ \\
PZSIC-r$^2$SCAN &  -77$^\text{[a]}$ \\
vSOSIC-r$^2$SCAN & {-93}\\
\hline
Exp.	&	0 to -40$^\text{[c]}$\\ 
\bottomrule
\end{tabular}

\footnotesize{\textsf{
[a] Reference \citenum{mishra2022coupling} where the values are calculated with an ECP approach for the PZSIC method.
[b] ECP was used in the calculation.
[c] Reference \citenum{willett1985magneto}.
}}

\end{table}

\subsection{Spin charges in square planar copper complexes}
As the final case study, we applied the vSOSIC method to the square planar copper molecule previously studied by Karanovich \textit{et al.}\cite{karanovich2021electronic}  They analyzed the electronic configurations for monoanionic [Cu(C$_{6}$H$_{4}$S$_2$)$_2$]$^{-}$ (Q1) and dianionic [Cu(C$_{6}$H$_{4}$S$_2$)$_2$]$^{2-}$ (Q2) Cu-based molecules. 
Similar magnetic square planar structures [Cu(C$_{14}$H$_{20}$S$_2$)$_2$]$^{z}$ ($z=2-, 1-, 0$) are synthesized experimentally.\cite{ray2005electronic} 
Due to its long spin-lattice relaxation times, [Cu(C$_{14}$H$_{20}$S$_2$)$_2$]$^{2-}$ complex is considered as a candidate for qubits in the area of quantum information science.
There has been a debate about the electronic structures of these square planar metal structures, and it has been 
suggested  that beyond-DFT methods such as multireference methods or potentially SIC methods may be required to study these complexes.\cite{fataftah2019metal,maiti2014one,ray2005electronic,karanovich2021electronic}
Use of popular functionals such as PBE or B3LYP results in incorrect  
electron delocalization for the copper 
d-electrons.
The Mulliken spin population at the Cu site with PBE is 0.32 $\mu_B$, compared to the EPR experimental value of 0.51 $\mu_B$.
PZSIC-LSDA yields 0.67 $\mu_B$, which is more similar to the Hartree-Fock and CASSCF estimates (0.79 and 0.70 $\mu_B$, respectively).
\cite{karanovich2021electronic}
The PZSIC tendency to overestimate the spin population has also been observed in 
spin-crossover complexes.\cite{romero2023spin}
Within the PBE and B3LYP functionals, the HOMO energy for the Q2 complex is positive, indicating that the additional 
electron is not bound to the complex. 
In both complexes, HOMO energy decreases from PBE to PZSIC-LSDA by 4.6 and 5.0 eV for Q1 and Q2 complexes, respectively.
The 
HOMO and LUMO eigenvalues and spin population obtained with vSOSIC are presented in Table \ref{tab:cu-based}. 
Both the PZSIC-LSDA and vSOSIC-LSDA HOMO energies are negative and agree within 0.1 eV.
Similarly, the difference in Mulliken spin population between PZSIC and vSOSIC is 0.12$\mu_B$ and 0.01$\mu_B$ for the Cu and S atoms in Q2.
The spin population for Q1 is excluded from the table since their spin moment is zero.
In this SOSIC calculation, 100 orbitals out of 175 total orbitals are treated with SIC reducing the computation time by 57\%. The reduced FOD structure for the vSOSIC method is shown in Fig. \ref{fig:cu2cl6_fod}.
PZSIC calculations on transition metal complexes surrounded by ligands are often computationally costly.  These systems demonstrate
how vSOSIC can be significantly more computationally efficient than PZSIC.

\begin{table*}
\caption{\label{tab:cu-based}
The Mulliken population on the Cu and all four S atoms and HOMO and LUMO eigenvalues of (Q1) [Cu(C$_{6}$H$_{4}$S$_2$)$_2$]$^{1-}$  and (Q2) [Cu(C$_{6}$H$_{4}$S$_2$)$_2$]$^{2-}$.
}
\begin{tabular}{@{\extracolsep{\fill}}lcccccc}
\toprule
\multirow{2}{*}{System} & \multirow{2}{*}{Method}	   & \multicolumn{2}{c}{Mulliken population ($\mu_B$)} & \multicolumn{1}{c}{HOMO}	&  \multicolumn{1}{c}{LUMO} & 	\multirow{2}{*}{No. SIC orbitals} \\\cmidrule(lr){3-4} 
	&&	Cu	&	S	&	(eV)	&	(eV)	  &	\\\hline
\multirow{2}{*}{Q1} & PZSIC-LSDA	&&&	-5.9	&  -1.2   &  174  \\
  &SOSIC-LSDA	&&&	-5.9	&  -1.4   &  100  \\
 \midrule
\multirow{2}{*}{Q2}&	PZSIC-LSDA	&	0.67	&	0.33	&	-2.0	&  4.4 &  175    \\
&	SOSIC-LSDA	&	0.55	&	0.34	&	-1.9	&  4.1 & 101  \\
    \bottomrule
    \end{tabular}
\end{table*}

\section{Efficiency of vSOSIC method}

 As a reminder, PZSIC is a one-electron SIC approach wherein 
 self-interaction correction is obtained in an orbital-by-orbital fashion. Thus,
as a rough estimate, the computational cost of PZSIC calculation is $N+1$ times that of a DFA 
(e.g. the standard LSDA) calculation, where $N$ is the number of electrons in the system. 
In practice, however, the PZSIC cost of calculation can be much higher since 
localized orbital densities need to be determined for evaluation of SIC.
In the FLOSIC method, 
this amounts to the optimization of FODs that determine the FLOs used to 
evaluate SIC terms. Our experience shows that the FOD energy surface is 
often very shallow with multiple minima, and typically the FODs corresponding
  to the core orbitals, especially the 1s orbital, are harder to optimize. Such
problems make the FLOSIC calculations time-consuming. Thus, by applying 
SIC to select orbitals, computational costs can be substantially reduced.

There are two ways in which vSOSIC accelerates SIC computations. 
Firstly, as done 
in this work, with only the valence orbitals, the vSOSIC calculation for a given set of FODs 
is 
$(N+1)/(N - N_{core}+1)$ times faster compared to a regular PZSIC calculation ($N-N_{core}+1$ times the standard LSDA). Here $N_{core}$  and $N$ are the number of core orbitals and total orbitals, respectively.
This also results in having to optimize 
$3(N - N_{core})$ FOD parameters in vSOSIC instead  of $3N$ parameters as in the PZSIC. 
As mentioned earlier, the optimization of the core FOD parameters is often
problematic and time-consuming. We find in this work that their removal (vSOSIC) usually 
leads to smoother optimization.

\begin{figure*}
    \centering
    \begin{subfigure}[b]{0.40\textwidth}
       \centering
       \includegraphics[width=0.9\textwidth]{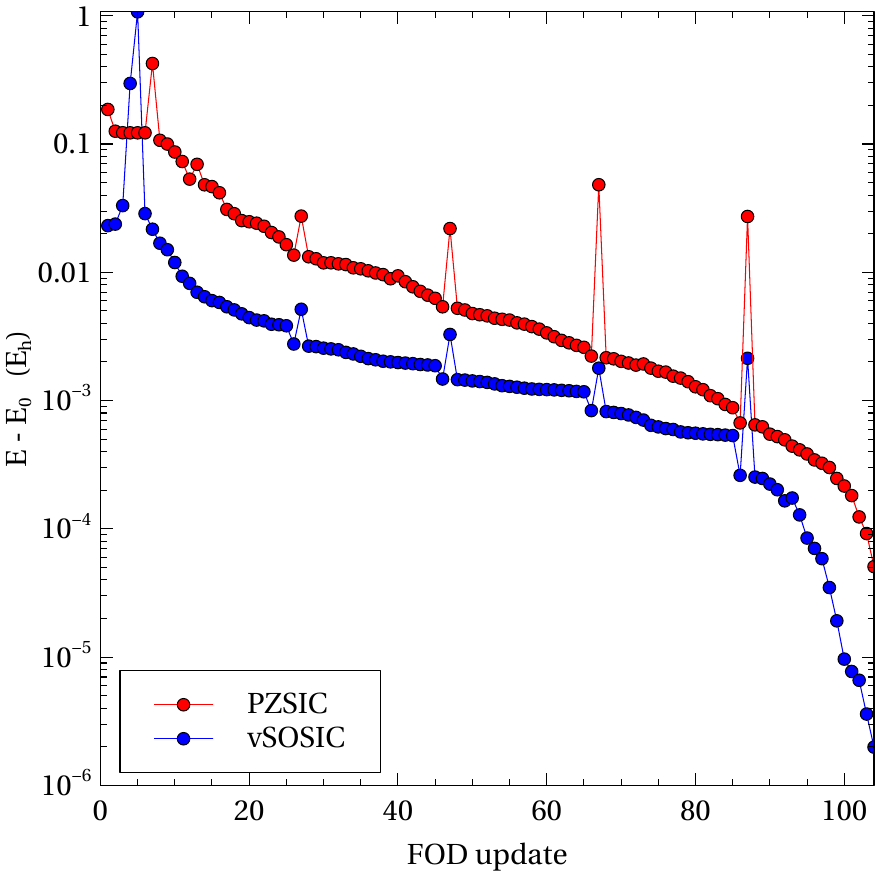}       
       \label{fig:ener-sosic-conv}
    \end{subfigure}
    \begin{subfigure}[b]{0.40\textwidth}
       \centering
       \includegraphics[width=0.9\textwidth]{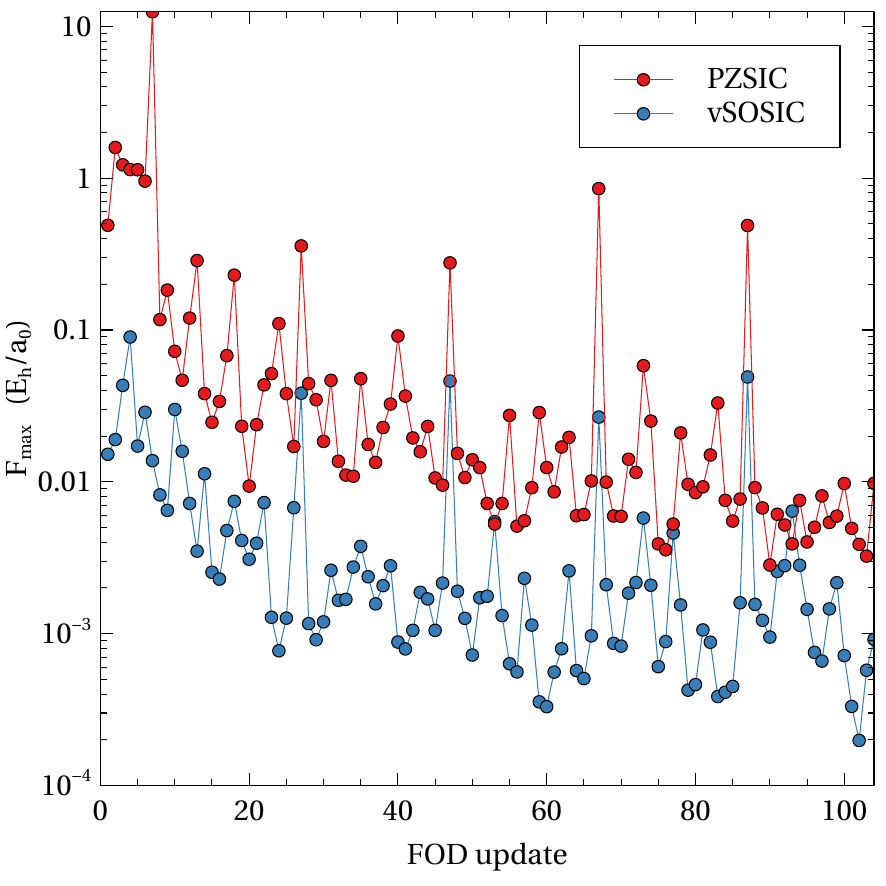}
       \label{fig:fod-sosic-conv}
    \end{subfigure}
    \caption{
    PZSIC and vSOSIC for the relative total energies with respect to their final converged energies as a function of the FOD update steps (left pane) and the largest FOD force component as a function of the FOD update (right pane).
   }
    \label{fig:fod_conv}
\end{figure*}

As an illustration of computational savings, we consider 
the [Cu$_2$Cl$_6$]$^{2-}$ complex and perform vSOSIC and PZSIC with LSDA functional calculations 
on the NERSC Perlmutter supercomputer that is equipped with AMD EPYC 7763 CPU with base (max) clock speed 2.45GHz (3.5GHz). 
We utilized two CPU sockets, resulting in a total of 128 cores per node. For conducting the timing test, only one node was used. 
The number of electrons treated with SIC in PZSIC (vSOSIC) is 160 (80), and a single iteration step takes 343.4 (174.6) seconds with the above setup. This speedup in an SCF from PZSIC to SOSIC is in agreement with the expected speedup of $(N+1)/(N-N_{core}+1)\approx 2$.
In Fig. \ref{fig:fod_conv} (a) we show the relative energy with respect to their converged energies as a function of FOD update steps for the same system.
The FOD optimization in PZSIC requires $\approx$105 steps to reach the final energy within $10^{-4}$ E$_\text{h}$  (where the tolerance in the energy derivative is $10^{-2}$ E$_\text{h}$/a$_0$).
In the vSOSIC, the relative energy, as well as the energy derivative, is almost one order lower than the 
PZSIC at the same FOD update step.

The largest component of FOD force is plotted as a function of update steps in Fig. \ref{fig:fod_conv} (b).  For this system, in 
 the PZSIC method, force drops below $10^{-2}$ E$_\text{h}$/a$_0$ after around 45 steps, whereas in the 
 vSOSIC  method, only 10 optimization steps are required to achieve the same, thus 
illustrating that the vSOSIC method being advantageous in FOD optimization.

\section{Summary and Outlook}\label{sec:discussion}

We have outlined and assessed a simplified one-electron self-interaction-correction scheme where
the SIE is removed from a select set of orbitals. 
Most often our interest is in the electronic  structure and chemical  bonding, making 
thereby the valence orbitals to be the orbitals of interest for the SIE removal as 
it is the valence orbitals that define central aspects of the electronic 
structure and chemical  bonding.  
The approach is, however, more general and can also
be used to correct for core states if needed, for example, in the computation of
core-electron binding energies (cf. supplementary information). 
We have illustrated the SOSIC approach  using both the valence and core as active set 
of orbitals for SIE removal.
The present  SOSIC approach can also be adapted to apply SIC to a specific region of space, as in the spirit of embedding methods, by identifying the FLOs that are localized in the region of interest.
We have shown that the vSOSIC or cSOSIC results in a substantial reduction in the computational complexities  by reducing the 
number of Fermi orbital descriptors that need to be optimized, thereby providing  
significant computational speedup. The results obtained using the vSOSIC and cSOSIC 
schemes  are 
compared with those obtained with PZSIC, which corrects for
all the orbitals. 
We have studied the performance of vSOSIC on the total energies of atoms, 
atomization energies, reaction barrier heights, and HOMO eigenvalues of molecules. 
For the atomization energies of AE6 datasets, with the LSDA functionals MAE 
with vSOSIC MAE is larger by 7 kcal/mol than the PZSIC, for the PBE functionals 
vSOSIC MAE is 2 kcal/mol lower while for
the r$^2$SCAN functional vSOSIC MAE is 
9 kcal/mol smaller than that of PZSIC. These differences between the performance 
of vSOSIC and PZSIC diminish for a larger dataset (37 molecules from the MGAE109).
 The 
vSOSIC and PZSIC perform similarly
within 0.2 kcal/mol in the calculation of the barrier heights of BH6 and WCPT18 datasets. Likewise, the absolute HOMO eigenvalues that approximate the 
vertical ionization energies,  obtained by the vSOSIC and PZSIC are in excellent 
agreement with each other but they both overestimate the experimental ionization 
energies.

\begin{sloppypar}
Furthermore, we applied vSOSIC to  
VDEs 
of water cluster anions 
and magnetic exchange coupling parameters of [Cu$_2$Cl$_6$]$^{2-}$ and 
the electronic structure of [Cu(C$_6$H$_4$S$_2$)$_2$]$^{1-/2-}$as a test 
on  systems containing transition metals. 
The vSOSIC and PZSIC predicted exchange coupling constants differ by 6 and 
10 cm$^{-1}$ for the LSDA and PBE functionals, respectively. 
For the [Cu(C$_6$H$_4$S$_2$)$_2$]$^{1-/2-}$ molecules, the vSOSIC, like PZSIC 
binds the extra electrons and yields HOMO eigenvalues within 0.1 eV of PZSIC, 
while for the spin moment at Cu site, vSOSIC prediction (0.55 $\mu_B$)
agrees with PZSIC (0.67 $\mu_B$) within 0.1 $\mu_B$ with vSOSIC value
being closer to the EPR experimental value of 0.51 $\mu_B$.
The water cluster anions offer an interesting  case.
Our previous work with PZSIC-PBE showed that the negative of the highest occupied 
eigenvalue offered an outstanding approximation to the VDE of the water cluster 
anions, with an MAE of only 17 meV when compared to CCSD(T) values. 
These results outperformed MP2 method and other hybrid functionals by 
a wide margin. The vSOSIC technique (with PBE functional) decreases the MAE by 
another 2 meV, making it an ideal alternative to CCSD(T) for determining the VDE of water cluster anions.
To determine the VDE of the water anions, an even more straightforward form of SOSIC was explored,
in which just the outermost unpaired orbital was corrected for SIE (1orb-SOSIC). Interestingly in this 
scheme, we find that the HOMO eigenvalue is negative indicating  electron binding. 
However, the VDEs derived from the HOMO eigenvalues, in this case, exhibit a higher MAE of 
56.9 meV, which is still superior to B3LYP (238 meV) and comparable to MP2 MAE (44 meV).
These 1orb-SOSIC results are highly encouraging because the computing cost is nearly the same as
that of the  uncorrected density functional technique making 1orb-SOSIC useful in molecular dynamics 
simulations of such complexes. 
\end{sloppypar}

The vSOSIC calculations on the  [Cu$_2$Cl$_6$]$^{2-}$ complex as an instructive example of 
the computational  efficiency demonstrates that, in addition to the savings from using 
fewer orbitals to account for SIC, the FOD optimization in vSOSIC is substantially smoother and quicker.
Overall, the vSOSIC method involves fewer calculations than the PZSIC method 
and produces results similar to the PZSIC method. As seen in the computation of the 
VDE of water anions, the approach may be tailored to the task at hand by selecting relevant orbitals for SIE removal.
For example, the properties related to the core orbitals such as core electron 
binding energies or Fermi-contact terms 
will require different choices of the active orbitals in the SOSIC method.
Interested readers may refer to the supplementary information for the SOSIC application to the core electron binding energies
where we included PZSIC and SOSIC performance for the case. 
The vSOSIC approach can be  particularly useful for studying a large complex composed of heavy elements where
SIC effects are expected to be more pronounced due to localized f-electrons. 

  Besides introducing and assessing the vSOSIC method, this work also assessed the performance of the r$^2$SCAN functional with PZSIC and vSOSIC 
  methods for a range of properties. Our results show that SIC-r$^2$SCAN calculations require 
  about 2-3 times fewer grid points than the SIC-SCAN calculations. SIC-r$^2$SCAN performs 
  similarly to the  SIC-SCAN for most properties but for atomization energies, 
  SIC-r$^2$SCAN outperforms SIC-SCAN.

\section*{Data Availability Statement}
The data that support the findings of this study are available within the article and its supplementary material.

\section*{Acknowledgement}
This work was supported by the U.S. Department of Energy, Office of Science, Office of Basic Energy Sciences, as part of the Computational Chemical Sciences Program under Award No. DE-SC0018331. Support for computational time at the Texas Advanced Computing Center (TACC), 
the Advanced Cyberinfrastructure Coordination Ecosystem: Services \& Support (ACCESS) program, 
and the National Energy Research Scientific Computing Center (NERSC) is gratefully acknowledged.

\section*{Conflict of Interest}
The authors declare no conflict of interest.

\begin{shaded}
\noindent\textsf{\textbf{Keywords:} \keywords} 
\end{shaded}


\setlength{\bibsep}{0.0cm}
\bibliographystyle{Wiley-chemistry}
\bibliography{bibtex}

\clearpage

\clearpage


\section*{Entry for the Table of Contents}

\noindent\rule{11cm}{2pt}
\begin{minipage}{5.5cm}
\includegraphics[width=5.5cm]{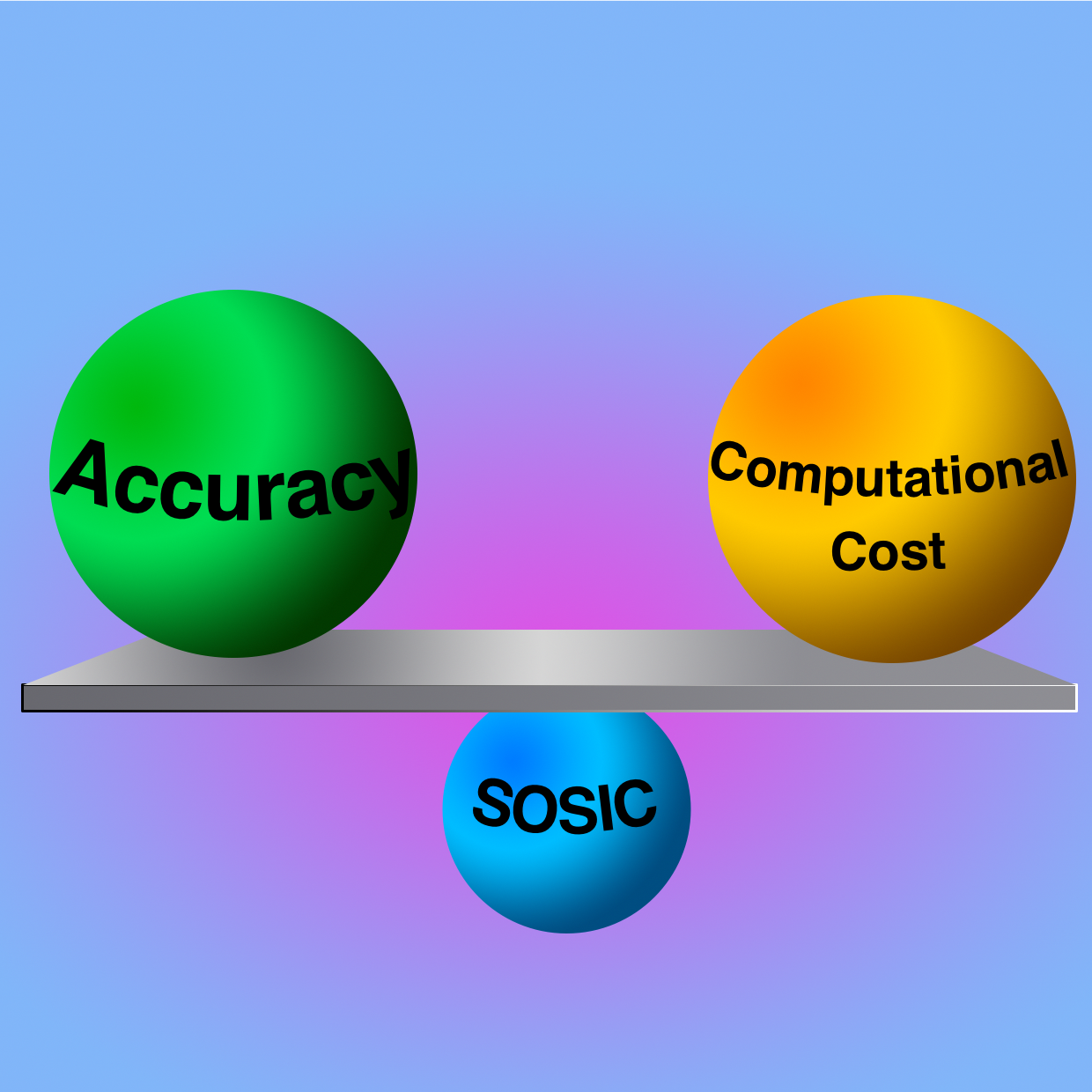} 
\end{minipage}
\begin{minipage}{5.5cm}
\large\textsf{
Many failures of density functional approximations are attributed to the self-interaction error. A simplified one-electron self-interaction correction (SIC) scheme is proposed to expedite the SIC calculations. Applications show that for a wide range of properties, it performs similarly to the well-known Perdew-Zunger SIC method at a substantially reduced computational cost.}
\end{minipage}
\noindent\rule{11cm}{2pt}

\vspace{2cm}

\end{document}